\documentclass[acmsmall]{acmart}
\AtBeginDocument{%
  }

\setcopyright{acmlicensed}
\acmJournal{PACMHCI}
\acmYear{2025} \acmVolume{9} \acmNumber{2} \acmArticle{CSCW054} \acmMonth{4}\acmDOI{10.1145/3710952}

\acmISBN{978-1-4503-XXXX-X/18/06}

\begin{document}

\title{Cultivating a Supportive Sphere}
\subtitle{Designing Technology to Increase Social Support for Foster-Involved Youth}
\author{Ila K Kumar}
\orcid{0000-0002-3482-6615}
\email{ilak@media.mit.edu}
\affiliation{%
  \institution{Massachusetts Institute of Technology}
  \streetaddress{75 Amherst St}
  \city{Cambridge}
  \state{MA}
  \country{USA}
  \postcode{02139}
}

\author{Craig Ferguson}
\email{fergusoc@media.mit.edu}
\affiliation{%
  \institution{Massachusetts Institute of Technology}
  \streetaddress{75 Amherst St}
  \city{Cambridge}
  \state{MA}
  \country{USA}
  \postcode{02139}
}

\author{Jiayi Wu}
\email{jw103@wellesley.edu}
\affiliation{%
  \institution{Wellesley College}
  \state{MA}
  \country{USA}
}

\author{Rosalind W Picard}
\email{picard@media.mit.edu}
\affiliation{%
  \institution{Massachusetts Institute of Technology}
  \streetaddress{75 Amherst St}
  \city{Cambridge}
  \state{MA}
  \country{USA}
  \postcode{02139}
}

\renewcommand{\shortauthors}{Ila K. Kumar, Craig Ferguson, Jiayi Wu, and Rosalind W. Picard}
\begin{abstract}
  Approximately 400,000 youth in the US are living in foster care due to experiences with abuse or neglect at home \cite{bureau_adoption_2023}. For multiple reasons, these youth often don’t receive adequate social support from those around them. Despite technology’s potential, very little work has explored how these tools can provide more support to foster-involved youth. To begin to fill this gap, we worked with current and former foster-involved youth to develop the first digital tool that aims to increase social support for this population, creating a novel system in which users complete reflective check-ins in an online community setting. We then conducted a pilot study with 15 current and former foster-involved youth, comparing the effect of using the app for two weeks to two weeks of no intervention. We collected qualitative and quantitative data, which demonstrated that this type of interface can provide youth with types of social support that are often not provided by foster care services and other digital interventions. The paper details the motivation behind the app, the trauma-informed design process, and insights gained from this initial evaluation study. Finally, the paper concludes with recommendations for designing digital tools that effectively provide social support to foster-involved youth.
\end{abstract}

\ccsdesc[300]{Human-centered computing~User studies}
\ccsdesc[500]{Human-centered computing~Usability testing}
\ccsdesc[500]{Human-centered computing~Social networking sites}
\ccsdesc[300]{Human-centered computing~Smartphones}
\ccsdesc[500]{Human-centered computing~Empirical studies in collaborative and social computing}

\keywords{Foster care, Transitional age youth, Foster-involved youth, Social support, Emotional support, Esteem support, Social network support, Peer-to-peer support platform, User research, User experience design, Qualitative research}

\received{January 2024}
\received[revised]{July 2024}
\received[accepted]{October 2024}

\maketitle

\section{Introduction}
At this moment, approximately 400,000 youth in the United States are living in temporary housing situations (known as foster care) because they have been identified as not receiving adequate care in their homes \cite{bureau_adoption_2023}. The vast majority of foster-involved youth (76\%) have been removed from their homes because their physical or emotional needs were not being met by their primary caregivers (i.e., the need for healthy food, a clean living environment, and emotional support) \cite{bureau_adoption_2023,health_administration_2021}. Despite the system's mission to fill the gaps in youths' care, foster care often separates youth from sources of support in their community and makes it difficult for them to form new helpful connections \cite{health_administration_2021, bureau_adoption_2023,benedict_types_1994, noauthor_national_2022, spencer_out--home_1992, tittle_child_2001,cancel_aged_2022, fathallah_away_2021,villagrana_foster_2021}. Indeed, approximately 40\% of these youth report that they do not have enough people to turn to for social support by the time they are teenagers \cite{okpych_memo_2018}. This lack of social support, encompassing direct help, guidance/advice, care/concern, encouragement, and a sense of belonging, negatively impacts youths' well-being and ability to navigate stressful life situations \cite{khan_social_2010,cohen_stress_1985,herman_trauma_2015,kawachi_social_2001}, with over 50\% of foster-involved youth diagnosed with a mental illness by the age of 19 \cite{courtney_mental_2015}, over 50\% involved with the juvenile legal system by the age of 17 \cite{courtney_midwest_2004}, and 31\% to 46\% experiencing homelessness by the age of 26 \cite{dworsky_homelessness_2013}. 

As internet-connected technology such as smartphones and computers have become more affordable and accessible, new opportunities are arising to increase social support for foster-involved
youth \cite{lehtimaki_evidence_2021}. Researchers believe that technology can help these youth build and strengthen supportive connections, find a community and explore their identity, and provide them with opportunities to support others \cite{brinson_becoming_2015, denby_promoting_2016, gustavsson_positive_2015}. Indeed, a recent  meta-analysis of eight peer-reviewed publications describing foster-involved youths’ experiences with internet-connected technology found that youth use technology to stay connected with family and friends (both those in their community and those that they no longer live close to), find resources, feel a sense of connection with peers who are not in foster care, and engage in self-care (through music, pictures, and games) \cite{sage_systematic_2021}. And in the only known study on the use of a social platform by individuals with foster care experience (analyzing a public reddit community for former foster youth), individuals with foster care experience used the platform to share personal information in ways found to be distinct from those who have not spent time in the foster care system, potentially to be heard and validated by others \cite{fowler_policy_2023}. 

At the same time, scholars and foster-involved youth worry that having access to internet-connected technology may increase youths’ exposure to cyberbullying, harassment, predatory solicitations, and other forms of abuse utilizing their personal information \cite{gustavsson_positive_2015}. A small-scale study conducted with foster parents found that more than half of foster-involved teens interacted with unsafe people online, leading to rape, sex trafficking, and/or psychological harm \cite{badillo-urquiola_risk_2019}. The majority of foster parents in this situation did not know how to ensure youths’ safety online and resorted to removing technology access altogether \cite{badillo-urquiola_risk_2019}. This may in part be because there has been almost no research done on designing safe technology interactions for foster-involved youth, with existing research on designing safe technological interactions for teens focusing heavily on parental mediation which is often not applicable to youth in foster care \cite{badillo-urquiola_understanding_2017}. In the few studies that explore youths’ perspectives on this topic, youth say they mitigate risks by being careful about what they share online (i.e., only communicating with individuals whom they know) and going to adults for help if needed \cite{sage_systematic_2021}. To tackle the unique tensions posed by technology for foster-involved youth, some scholars advocate for more research focusing on how to design safe and helpful technology for these youth, prioritizing their perspectives \cite{badillo-urquiola_understanding_2017, wisniewski_using_2017}.

The lack of adequate support for foster-involved youth and gaps in literature on how technology can be leveraged for these youth have motivated us to explore how internet-connected technology (particularly online platforms) can be used to increase social support for this population. Towards this end, we designed, developed, and evaluated a mobile application, Sphere, aimed specifically at increasing social support for foster-involved youth. In the evaluation process, we focused on the research question: How can foster-involved youth gain and provide social support in online platforms, and what design features are responsible for fostering social support in these settings? Across our research methodologies and design decisions, we prioritized human-centered, trauma-informed, and healing-centered approaches to ensure that the project’s process and output were guided by the safety, needs, and perspectives of foster-involved youth \cite{dietkus_call_2022, ginwright_future_2018, soto-aponte_crosswalk_2021, noauthor_substance_2014}. In this paper, we describe the motivation for the app, the collaborative and trauma-informed process that led to its design, and the takeaways from an initial evaluation study conducted with current and former foster-involved youth. We end with our main contribution: recommendations for how to design digital tools that provide social support to foster-involved youth. 

\section{Related works}

\subsection{Social support components and impact on well-being} \label{relatedworks_socialsupport}

Social support can be broken down into tangible support (providing direct help), informational support (providing knowledge or advice), emotional support (expressing care, concern, and empathy), esteem support (boosting one’s sense of ability and value), and social network support (having a sense of belonging to a group with similar interests or experiences) \cite{ko_understanding_2013, huh-yoo_help_2023}. Social support has been shown to be beneficial for improving well-being, with multiple studies highlighting that social support increases positive psychological strengths such as hope, optimism, self-efficacy, and resiliency \cite{khan_social_2010}, directly reduces the severity of negative psychological symptoms, and protects individuals from the potentially adverse effects of stressful situations \cite{cohen_stress_1985,herman_trauma_2015,kawachi_social_2001}. Research with foster-involved youth has also shown that youths’ well-being is improved when they are provided with emotional support, information, and opportunities for self-reflection \cite{munson_steady_2010,munson_natural_2009}.

\subsection{Systemic lack of social support for foster youth}

Existing services do not adequately fill the gap in social support for foster-involved youth. Being placed away from siblings and others in their community and experiencing frequent moves makes it difficult for many youth to naturally build and maintain supportive relationships with others in their lives \cite{children_adoption_2021, bureau_adoption_2023, hegar_foster_2011, noauthor_how_2020, mccormick_siblings_2010}. While youth are meant to be overseen by one or more caregivers throughout their time in foster care, research suggests that many foster-involved youth experience neglect or abuse at the hands of these individuals \cite{benedict_types_1994, noauthor_national_2022, spencer_out--home_1992, tittle_child_2001}. Foster-involved youth are also assigned a social worker who is employed to help them receive the services and resources they need while in foster care. However, youth report feeling that social workers do not always know or care about the challenges they are facing, and can be difficult to reach when they need help \cite{cancel_aged_2022, fathallah_away_2021}. Additionally, even though many foster-involved youth are offered mental health services like individual and group therapy while in care \cite{roos_therapist_2013}, youth report feeling that mental health services are ineffective and do not understand why they are being forced to attend, in part because these services are often mandated by the court \cite{villagrana_foster_2021}. This sentiment is believed to be why usage of mental health services drops by 50-60\% as soon as youth transition out of care \cite{villagrana_racialethnic_2017,villagrana_perceived_2018}. Foster-involved youth are sometimes able to participate in mentorship programs, in which they are paired with an unrelated adult for a set period. A recent meta-analysis indicates that these programs have a small-to-moderate effect on youths’ educational achievement and psychological state (e.g., feelings of self-determination and hope, traumatic stress symptoms, and anxiety and depression) \cite{poon_meta-analysis_2021}. Studies also suggest that youth who experience emotional abuse are less satisfied with and gain less from mentorship programs, potentially because they have more difficulty forming trusting relationships \cite{blakemore_impacts_2017,poon_meta-analysis_2021,ku_natural_2021,wilson_interpersonal_2015}. Any support youth receive through the foster care system tends to fall away once they transition out of care, leaving them to navigate life challenges alone \cite{cancel_aged_2022}. And although foster-involved youth usually have the option to stay in care up to age 21 (or 23 in some states) \cite{gateway_extension_2022,john_chafee_nodate}, approximately 75\% choose not to participate, with some citing that they want more freedom or do not want to continue dealing with social workers \cite{courtney_findings_2014,foundation_fostering_2018}.

\subsection{Limitations of existing technologies for foster-involved youth social support}

While some internet-connected technologies have been created to support foster-involved youth, none have been shown to provide holistic social support. In particular, there has been a lack of interventions that increase emotional, esteem, or social network support for foster-involved youth. A review of the literature and publicly available applications revealed that the majority of tools designed for current and/or former foster-involved youth focus on informational support, creating databases of resources and advice for youth (e.g., FosterClub, Think of Us’s Virtual Support Services, Youth Matters: Philly, Know Before You Go, Our Community L.A.’s WIN program, and FOCUS on Foster Families). Social media platforms may be able to provide more types of social support by connecting foster care-experienced youth with others in their communities (e.g., reddit channels and Facebook groups) \cite{fowler_policy_2023}, but these are often not spaces that feel safe, leaving youth vulnerable to predators or harassment \cite{badillo-urquiola_risk_2019,fowler_fostering_2022,gustavsson_positive_2015,sage_systematic_2021}. A few online community platforms have been created specifically for the foster care community (e.g. FosterClub and Stepping Forward LA), but these either do not separate youth from providers or have more so been used to share resources with youth users, which may not foster environments for youth to share emotionally-charged content (often a prerequisite for receiving emotional or esteem support in response). In order to design solutions that are actually safe and effective, we believe technologists need to center the voices of current and former foster-involved youth and prioritize decisions that encourage emotional, esteem and social network support \cite{badillo-urquiola_understanding_2017,wisniewski_using_2017}.

\subsection{Potential of technology to increase social support for foster-involved youth}

Despite limited research with foster-involved youth, evidence suggests that internet-connected technologies have the potential to foster emotional, esteem, and social network support in other contexts. In particular, research suggests that online platforms that allow users to connect with others who have similar experiences may be able to help individuals feel more supported, particularly if these sites encourage self-disclosure and prioritize safety. 

\subsubsection{Shared identity and experiences}
Online platforms have the potential to increase social support for individuals facing a variety of challenges, particularly when they connect similar users. Amongst individuals living with mental illness, online platforms for individuals who have faced similar mental health challenges have been shown to increase social connectedness and sense of belonging, or social network support \cite{prescott_online_2020, naslund_future_2016}. For people experiencing major life transitions, using online networks that connect them with others dealing with the same topic, separately from their existing social networks, appears to provide informational, social network and emotional support \cite{zhang_separate_2022}. Additionally, multiple studies focused on matching users within diverse online social platforms have found shared identity (including age, gender, race, and interests) are important for facilitating socially supportive interactions, particularly for users with marginalized identities \cite{andalibi_considerations_2021, saksono_evaluating_2023, fang_matching_2022}. 


At the same time, connecting similar individuals together online may not always be supportive, as suggested by literature on online platforms for pregnancy and birth loss. While participants in an online support group for women of color who had experienced stillbirth or early infant loss found it valuable both for giving and receiving social support, they also spoke about how it sometimes increased negative emotions by reminding them of their loss experiences \cite{gold_online_2022}. In an interview-based study of people who had experienced pregnancy loss, online platforms had the potential to be supportive or unsupportive based on how individuals compared themselves with others, in part mediated by the perceived level of similarity in identity and past experiences \cite{barta_similar_2023}. 

\subsubsection{Anonymous self-disclosure}
Sharing personal experiences appears to play a significant role in the amount of social support individuals receive online. In a study focused on a postpartum depression subreddit, telling personal stories in posts helped users both seek and provide social support \cite{progga_just_2023}. From studies examining public reddit posts on mental health subreddits \cite{choudhury_mental_2014} and young people’s private Instagram messages \cite{huh-yoo_help_2023}, self-disclosure from both the help-seeker and helper, particularly authentic and negative self-disclosure, facilitates socially supportive interactions online. Studies examining the discussion of sensitive topics in public reddit communities (such as mental health and sexual abuse) suggest that anonymity is an important mediator of self-disclosure, allowing for more open sharing both in support-seeking posts and supportive comments \cite{andalibi_social_2018,choudhury_mental_2014}. 

Even if self-disclosure does not reveal personally identifiable information, there are some risks that may counteract its socially supportive benefits. As highlighted by individuals who had previously shared weight-related social media content, reflecting on past posts had the potential to make them feel more supported and validated, but also to trigger unwanted memories or thoughts \cite{karizat_i_2023}. In another study with first-generation, low-income students, disclosing this identity on social media platforms could provide them with increased social support, but also could expose them to prejudiced comments that decreased how supported they felt \cite{pyle_social_2023}. 

\subsubsection{Safety}
As discussed in the introduction, foster youth face significant risks online, including potential for experiencing bullying, harassment, rape, or sex-trafficking \cite{gustavsson_positive_2015, badillo-urquiola_risk_2019}. While the current approach is to restrict access altogether \cite{badillo-urquiola_risk_2019} and reactively support victims \cite{badillo-urquiola_towards_2024}, some scholars argue that it is important to consider new solutions that proactively promote online safety \cite{badillo-urquiola_towards_2024}. Outside of the child welfare sphere, researchers suggest rather than focusing on restricting teens’ usage of online platforms, teens should be given scaffolded opportunities to practice online self-disclosure and learn healthy privacy protection practices \cite{jia_risk-taking_2015, wisniewski_preventative_2015}. In research centering teen voice in designing safe online platforms, teen participants advocated for features that make it easier for users to customize the visibility of their content \cite{jean_baptiste_teens_2023}, as well as features that focus on preventing negative actions – prompting perpetrators to rethink their actions, blocking potentially harmful actions, and penalizing perpetrators to prevent multiple harms \cite{agha_strike_2023}. 

In this study, we take lessons from this body of related work, along with feedback from current and former foster-involved youth, to design and evaluate a novel system specifically focused on providing social support to foster-involved youth. We end by contextualizing our findings within existing knowledge about designing socially supportive online spaces and presenting design recommendations specific to the foster youth community. 

\section{App design}

\subsection{Community relationship building}

To responsibly engage with foster-involved youth, we partnered with five organizations that support young people who have experienced foster care in the United States. This was especially important for the project given that no members of the research team had experience being in foster care as a teenager. We felt it was important to center the perspectives of current and former foster youth at every stage of the project, as this population is frequently subject to programs and policies that do not reflect their needs, and there are many benefits to both the work and to the individuals shaping it, when they are invited to fully contribute \cite{salazar_authentically_2021}. For all phases of the project, we recruited participants exclusively through our partner organizations, with staff sending engagement opportunities to any eligible young people they were connected with. Guided by the principles of trauma-informed design, we held regular meetings with our partner organizations to build trust and transparency and gather feedback throughout the research process \cite{dietkus_call_2022}. Inspired by Healing Centered Engagement’s focus on encouraging youth to think about their own definitions of success and well-being, we also offered all of our community partners the opportunity to have their youth participate in workshops in which they explored what is possible in the field of technology and well-being, encouraging them to creatively imagine new technologies to support the needs of people like them \cite{ginwright_future_2018,soto-aponte_crosswalk_2021}. 

\subsection{App co-design process}

We spent eight months working alongside current and former foster-involved youth to identify core challenges and iterate on the platform concept and interface that became Sphere. We began by conducting seventeen virtual semi-structured interviews with current and former foster-involved youth to better understand the challenges youth face, the systems of support that already exist around them, and youths’ perspectives on how technology could be designed to meet their needs. Given no prior studies asked foster-involved youth for their views on how technology could support their well-being, we felt it was important to start with this question before prototyping. 

All procedures, materials, and data management systems were reviewed and approved by our institution's Institutional Review Board prior to implementation. Participants were given the option to attend the interview alone or with their case manager or mentor present (if applicable) based on their comfort levels and access to technology. At the start of each interview, we emphasized that participants were not expected to share any information that they did not wish to share. We also reminded participants that they were free to take a break, leave, or turn off their cameras at any time during the interview, for any reason. The interviewer made sure to keep their camera on regardless of whether the participant had their camera on so participants could read their body language, to increase transparency and make participants feel more comfortable. During the interview, we purposely posed general questions 
to avoid pressuring participants to share personal, potentially-traumatic information. Furthermore, we fostered a collaborative relationship with participants by treating them as thought partners throughout the interviews, asking them directly about what they felt would make a desirable and impactful technology output. 

In this first phase of interviews, participants primarily advocated for interventions that allowed them to receive support from and give support to others who have similar lived experiences (particularly with foster care). Participants also spoke about the importance of anonymity, community norms, and monitoring (ideally by an automated system to avoid a sense of surveillance) to ensure a safe and supportive environment. Because social support surfaced as a key need in these interviews, we subsequently focused the second phase of interviews on designing an application that might increase social support for youth in care. 

We then conducted twenty-four virtual prototype-testing interviews with current and former foster-involved youth, using methods that mirrored those of the first set of interviews. In the recruitment process, we prioritized transparency by trying to engage as many of the previous participants as possible, to share what we implemented based on their previous input and get feedback on whether we had left out anything that they felt was important. To avoid biasing participants, we asked for their ideas before showing them prototypes. However, we often found that participants found it easiest to contribute by critiquing and building off of prototypes, perhaps because it was less intimidating or sparked more of their own ideas. To encourage collaboration and honesty, we prefaced the sharing of any prototype by telling participants that they were being shown a very rough draft, that the team was eager to hear what should change to make the idea more useful, and that there would be no right or wrong answers to our questions. These prototype designs included a traditional forum (with topic channels, the ability to write posts, and the ability to react and respond to others’ posts), a peer-to-peer counseling model (where users optionally complete peer support training and then have the ability to provide written support to others), a Q\&A forum (exclusively for writing and responding to questions), and a reflective check-in platform. Through the latter, we explored whether participants preferred to share freely or in a structured high-low format, in addition to being able to react and respond to others’ check-ins. 

Based on the results of this process, we decided to move forward with a reflective check-in platform. In particular, this was the only platform in which participants expressed that they saw a clear way to receive and give emotional support, and sharing personal experiences, especially challenges, is a key factor in whether emotional, esteem and social network support take place online \cite{progga_just_2023, choudhury_mental_2014,huh-yoo_help_2023}. We decided to require users to share both a high and a low from their day in order to encourage negative self-disclosure (as this is particularly important for social support) while providing balance with something positive or hopeful. The high or low could be something that happened to them or something they were worried about or happily anticipating, as some examples. In case someone experienced negative emotions in the process of reflecting on their day or reading others' posts, we created a continuously accessible resource page where they could get access to immediate, 24/7 professional support. We also incorporated the ability to add questions/messages for the community to check-ins, as participants felt that this would allow them to share informational support (asking for or sharing resources) and esteem support (broadcasted messages of encouragement). We decided that the platform should be exclusively for individuals with verified lived experience in the child welfare system (including in formal settings like foster homes and group homes and informal settings like kinship care), as participants said this was important for social network support (feeling less alone in their experiences within or coming out of the child welfare system). This also aligned with prior research findings that online spaces which bring together individuals with common identities and experiences is important for facilitating social support, particularly social network support \cite{prescott_online_2020, naslund_future_2016,zhang_separate_2022,andalibi_considerations_2021,saksono_evaluating_2023,fang_matching_2022}. Based on the recommendations of work focused on safety in online interactions \cite{badillo-urquiola_risk_2019, badillo-urquiola_towards_2024,agha_strike_2023}, we also integrated features that aimed to proactively promote safety through anonymity and immediate removal of potentially triggering or harmful content. The anonymity feature was also intended to help youth feel more comfortable sharing their experiences with others, based in the findings of prior works \cite{andalibi_social_2018,choudhury_mental_2014}. 

Finally, we worked to refine the app through five virtual design workshops with four former foster-involved young adults. The first workshop focused on introducing participants to the research team and the project, and the following four workshops focused on iterating on the pilot app design. Each workshop began with a check-in in which all participants shared a high and low of their past week. This check-in activity helped the group (and particularly the facilitators) be sensitive to one another’s mood and provide support when needed. The facilitators participated in the check-in activity as well to model vulnerability and create a sense of equality amongst the group members (although we acknowledge that there were still power imbalances inherent to the team structure). After the check-in, facilitators presented core features of the app and gave participants opportunities to provide feedback, first written and then verbally via group discussion. In this phase, we made adjustments to the visual design, the reactions users could add to others’ posts, the feeling options users could select for check-ins, and the tags users could assign to their profile and use to filter content in their feed. 

After this, we transitioned to the focus of this paper, conducting a pilot study with a functional version of the app to evaluate the design's value for current and former foster-involved youth. 

\subsection{App design and technical specifications}

\begin{figure}[t]
    \centering
    \begin{minipage}[b]{0.25\textwidth}
        \includegraphics[width=\textwidth]{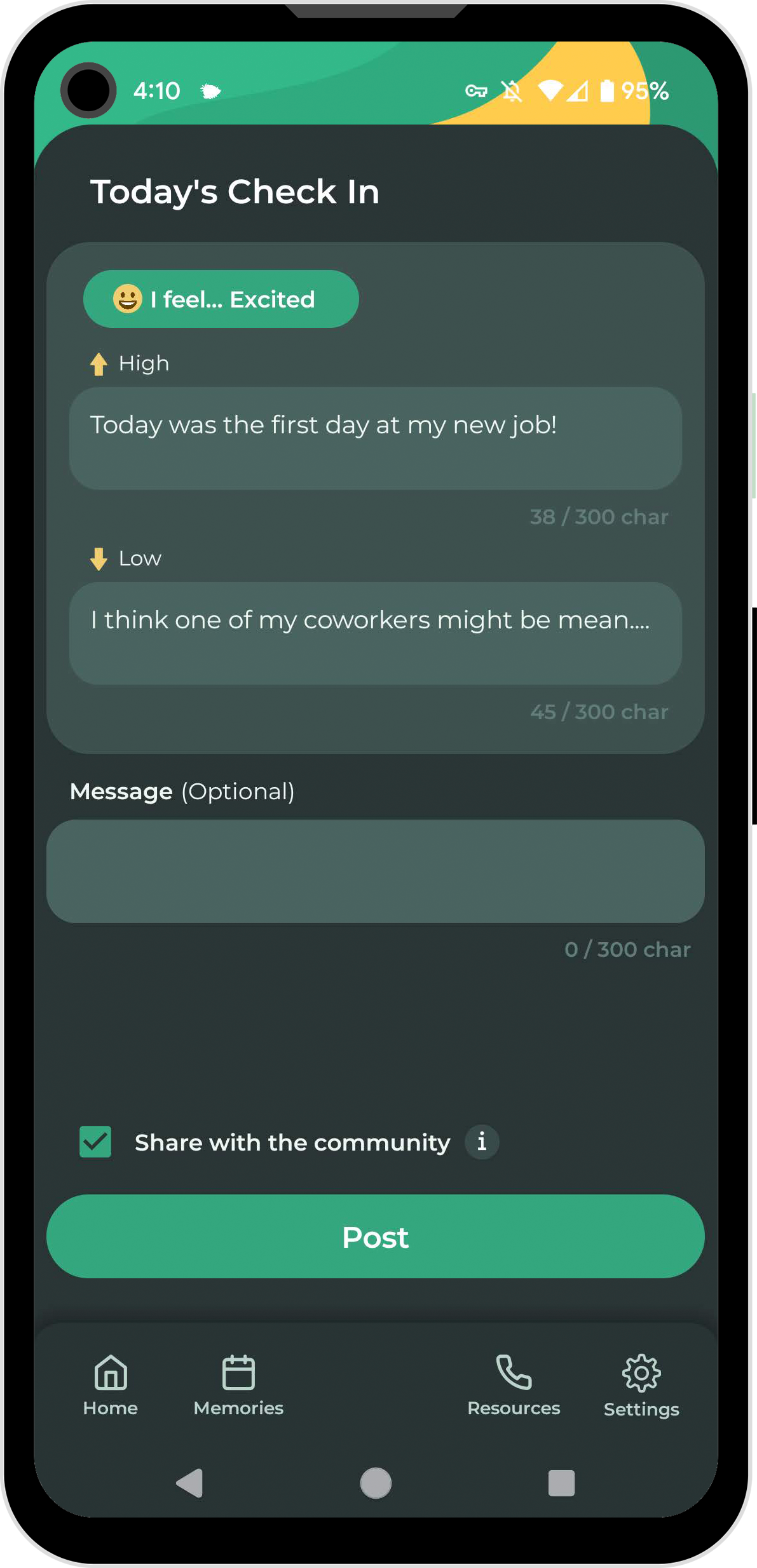}
        \caption{Check-in page}
        \Description{A depiction of a mobile phone screen with a header that says "Check in now". Below this is a user input form that begins with a button labeled "I feel...", and free-response boxes for users to fill in a "High", "Low", and "Message". Below this is a checkbox labeled "Share with the community" and finally, a "Post" button.}
        \label{fig:checkin}
    \end{minipage}
    \hfill
    \begin{minipage}[b]{0.25\textwidth}
        \includegraphics[width=\textwidth]{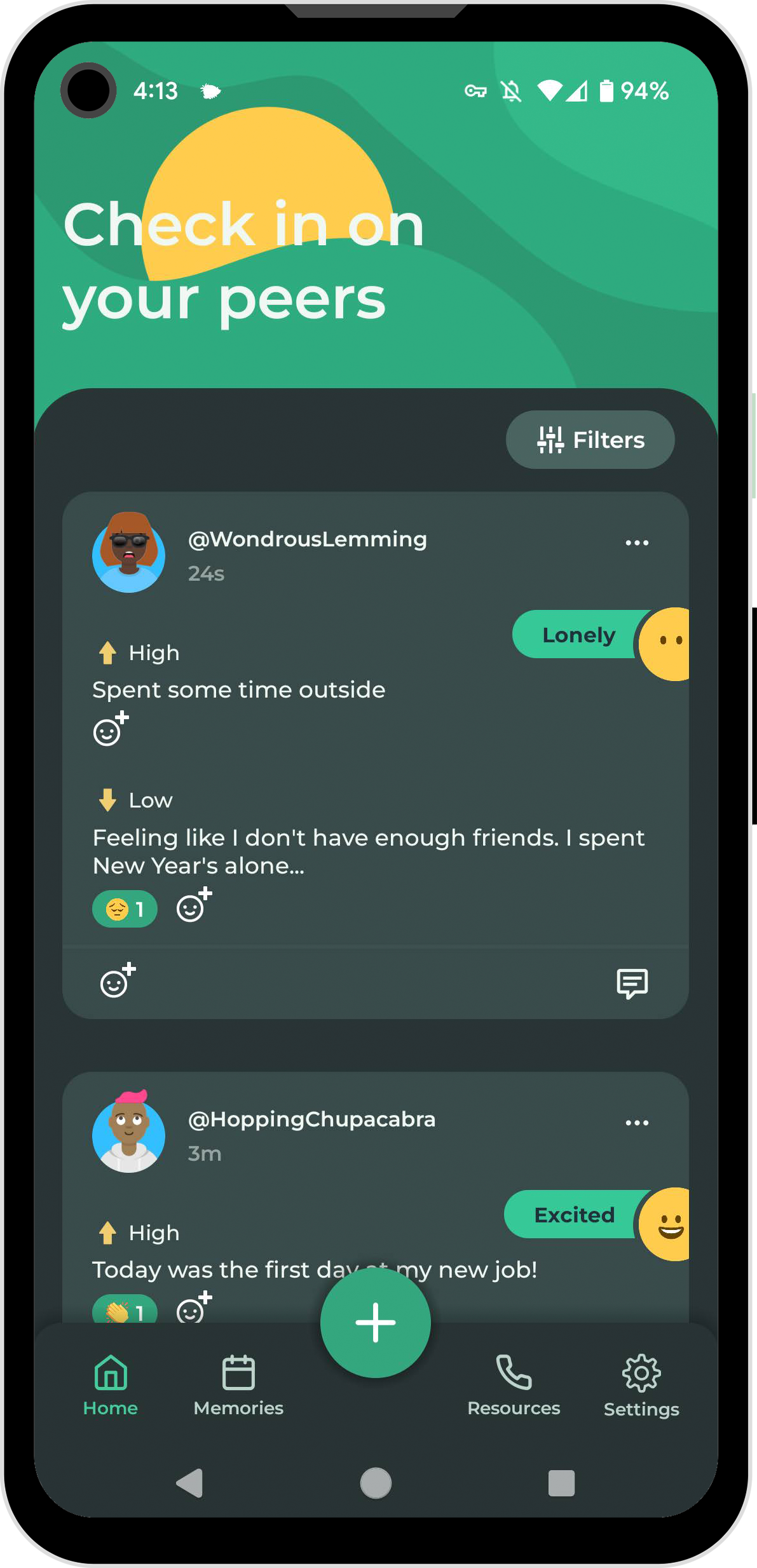}
        \caption{Community feed}
        \Description{A depiction of a mobile phone screen with a header that says "Check in on your peers". Below this is a filterable feed of posts from other users.}
        \label{fig:feed}
    \end{minipage}
    \hfill
    \begin{minipage}[b]{0.25\textwidth}
        \includegraphics[width=\textwidth]{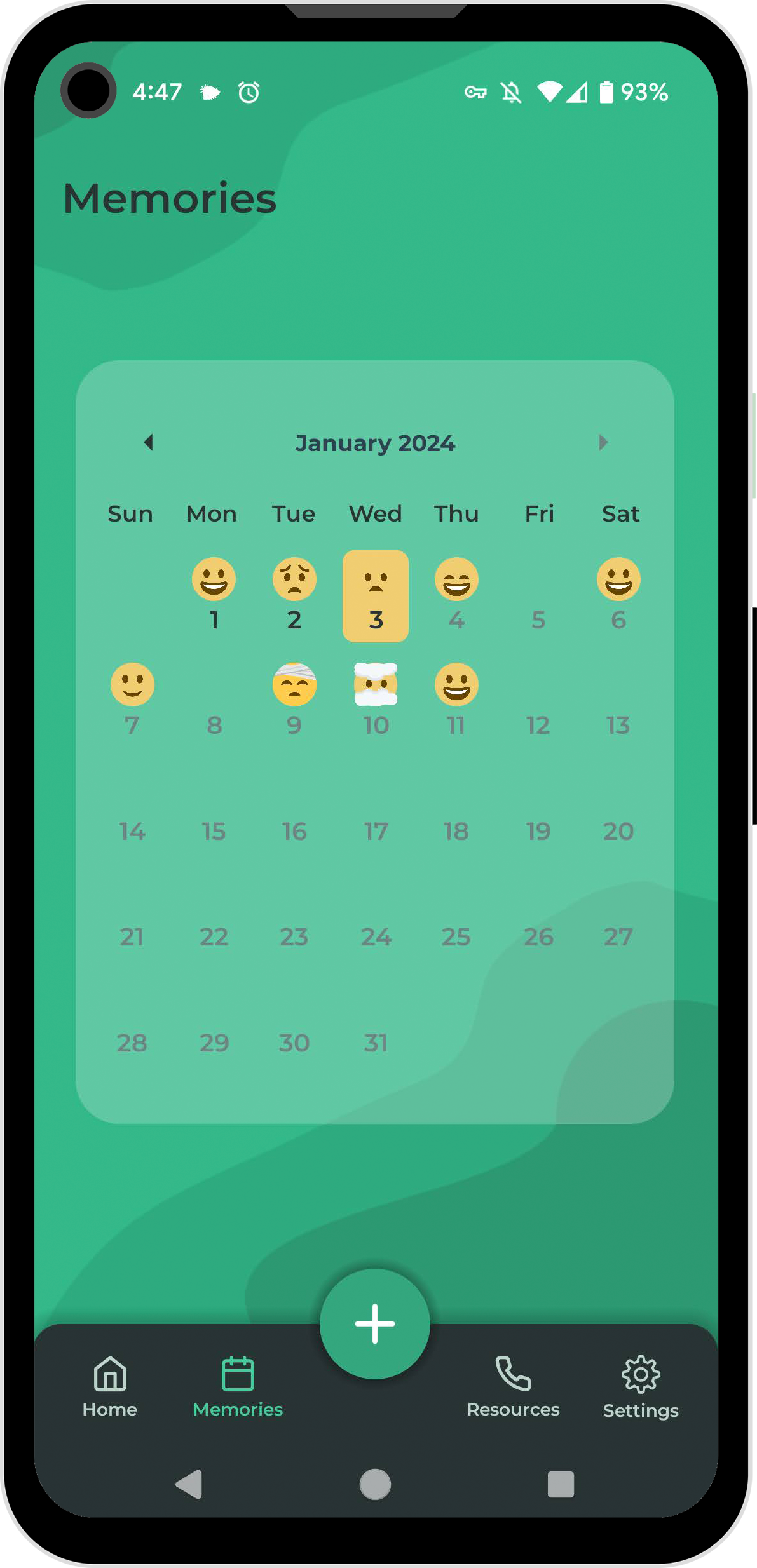}
        \caption{Memories page}
        \Description{A depiction of a mobile phone screen with a header that says "Memories". Below this is a calendar with a subset of days showing an emoji representing how the user was feeling on that day.}
        \label{fig:memories}
    \end{minipage}
\end{figure}

We created a cross-platform app named Sphere using React Native that logs data in a Django backend and stores it in a MySQL database. We deployed the Sphere app on the Google Play Store and Apple App Store. The app is free to download but requires a pre-made user ID to log in, which is assigned to users by their study coordinator. The app never logs any identifying information about the participant, using the assigned user ID as its only identifier. 

After downloading the app and entering their user ID, the app guides users through a process to set up their profile. It auto-generates random fun usernames consisting of a positive adjective followed by an animal name (e.g., “HappyHippo”). Although users can regenerate the name as many times as they want, they cannot manually pick their username to prevent users from picking negative or offensive usernames. The users are asked for some optional basic demographic information the app can use to sort/filter posts (relationship to foster care, education status, job status), and asks for the user to create a short bio that can be seen on their profile page. The app is careful to remind users to keep bios anonymous, and every bio is reviewed by study coordinators. Finally, the user can create an avatar with custom skin/hair/clothing combinations, to allow them to anonymously express themselves. Options range from realistic to fantastical, to let people represent themselves as they see fit.

Once the profile is complete, users are asked to agree to three community guidelines: 1) Be kind (no name-calling, trolling, threats, or insults), 2) Be safe (no asking for or sharing identifiable information; no sexually suggestive content; no discussion of illegal activities), 3) Prioritize comfort (no discussion of suicide, self-harm, abuse, or disordered eating; no cursing). They can then access the main screens of the app - the check-in page, the community feed, and the memories page. The check-in page (Figure \ref{fig:checkin}) allows users to complete a daily reflective check-in, which includes a high (something positive that happened recently), a low (something they’re struggling with), and an optional question or message (if they want to ask questions or share esteem support). It also asks users to select an emotion to represent how they’re feeling in the current moment. The emotion selection interface is based on the “feeling wheel” exercise used in Skills Training in Affective and Interpersonal Regulation (STAIR) Narrative Therapy to help trauma-exposed individuals learn to identify and label their feelings \cite{cloitre_stair_2022}. Users are limited to checking in on the app once a day, and if they revisit the check-in page after having reached this limit, a message invites them to visit the community feed to read other users’ check-ins. Users are given the option to share their check-in with the community (in the community feed, Figure \ref{fig:feed}) or keep it private (only visible on their memories page, Figure \ref{fig:memories}).

Users can see others’ public check-ins in the community feed, where they are able to write replies or react with emojis, which theoretically give users an easy way to provide support to others if they do not feel comfortable writing replies. Users can filter the community feed by basic demographic information (set during onboarding and updated through account settings) so that they can find posts from others who are similar to them.

Users can see all their past check-ins on the memories page, which shows all check-ins in a calendar view. At the month level, the calendar represents each past check-in by the feeling emoji selected on that day. If a user clicks on an emoji, it opens the full view of that check-in. When users receive new replies or reactions on their check-ins a red dot appears next to the associated feeling emoji, to indicate unread activity. Users also receive push notifications for any new replies or reactions on their check-ins, in addition to a daily notification at 6 p.m. local time reminding them to check in on the app if they haven’t already. Users can turn off any of these notifications on the settings page. 

All posts and replies are reviewed by machine learning models that auto-report potentially inappropriate content to moderators. Users also have the ability to report posts, replies, and user bios that violate community guidelines. If content is reported (either by the automoderation system or by other users), it is immediately hidden from all users (aside from the original poster) while moderators review it within 24 hours. The post is only made visible again if it is cleared by a moderator. This aims to minimize user distress by hiding harmful or triggering posts immediately. A banner appears above reported posts alerting the original poster of its status (under review or removed). Moderators can also leave private comments (only visible to the original poster) to let users know why their post has been removed or to provide resources (in the event that it is about potentially triggering content). To support high-need users who do not post about triggering content (and therefore may not be flagged by the moderation system), there is also a continuously accessible resource page that links to external platforms where users can find resources or get immediate crisis support via call or text (including Findhelp.org, TeenTalk, Crisis Text Line). 

In order to help with moderation, we developed a moderation tool platform based on Django’s admin framework. The moderation tools website (“modtools”) allows our study coordinators to log in and immediately see a high-level report of how many posts, replies, or user bios have been reported by users, auto-reported by the automoderation system, or removed by moderators. From there, modtools provides access to buttons that let the moderators quickly remove offending posts/replies/bios or reinstate ones that had been wrongfully reported. Additionally, all reports and moderator actions are logged, to see if any users abuse the report system and to monitor which moderation tools were required during the study.



\section{Methods}

\subsection{Data collection}

All procedures, materials, and data management systems were reviewed and approved by our institution's Institutional Review Board prior to implementation. In order to participate in the study, individuals needed to be proficient in English, between the ages of 16 and 24, and currently or previously involved in the child welfare system in the United States. Finally, participants needed to have continuous access to an iPhone or Android device for the two weeks in which they would be asked to use the app. 

At the start of the study (Touchpoint 1), interested individuals were asked to fill out the consent and baseline questionnaire, which contained a consent form, demographic questions, and the Short Self-Reflection Scale and UBC Social Connection Scale \cite{lok_ubc_2022,silvia_self-reflection_2022}. 

Approximately two weeks later (Touchpoint 2), participants were asked to complete a questionnaire consisting of the Short Self-Reflection Scale and the UBC Social Connection Scale. Then, participants were asked to download the app on their smartphones and create an account using a provided unique ID. After participants completed the onboarding process, they were reminded that in order to be compensated, they were required to complete one check-in a day on the app (allowing for some missed days). Beyond this, the research team emphasized that they could use the platform as much or as little as they liked. 

For approximately two weeks between Touchpoint 2 and Touchpoint 3, participants asynchronously used the app. During this time, the research team reviewed any check-ins or replies flagged by the content moderation system. Additionally, the team seeded the platform with five check-in posts created in advance by former foster-involved young adults and added emoji reactions to check-ins created by participants (at most one per post, and no replies) in order to simulate a more populated platform and better assess the impact of the social components of the platform. 

At Touchpoint 3, participants were asked to fill out the post-intervention questionnaire, beginning with the Short Self-Reflection Scale and UBC Social Connection Scale. Next, participants were asked a series of multiple-choice and short-answer questions designed to gather information about their experiences using the platform. This included custom questions catered to the study’s context and a modified version of the System Usability Scale \cite{brooke_sus_1996}. Finally, the questionnaire asked whether participants were interested in participating in a 30-minute follow-up video call to share more feedback about their experience for additional compensation. 

For those who chose to attend a 30-minute semi-structured follow-up video interview, questions focused on participants’ impressions of the app (e.g., what they liked, disliked, or found confusing) and their ideas for how the app could be improved. We also asked a general question about what kinds of technology they felt would be helpful for supporting youth who are in foster care, to investigate whether there was another direction they felt we should explore in the future. The methods of the interviews mirrored that of the app co-design interviews, except that participants were asked to join the call alone for consistency. 

We gave \$100 Amazon, Uber, or Apple gift cards (participant’s choice) to participants who completed all three questionnaires and used the app for two weeks. We gave participants an additional \$20 gift card if they participated in the optional interview.

\subsection{Data analysis}

App usage data were collected through the digital platform and survey data through Qualtrics. Interviews were conducted, recorded, and transcribed using Zoom. A deductive coding approach was used to analyze the interview transcripts, free-response survey answers, and app check-in posts and replies \cite{fereday_demonstrating_2006}. Five predefined codes (tangible support, informational support, emotional support, social network support, and esteem support) were derived from a framework of social support \cite{ko_understanding_2013}, as defined in Section \ref{relatedworks_socialsupport}. To ensure consistency and accuracy of code assignments, two members of the research team collaboratively analyzed a subset of the data using affinity diagramming techniques. The team began by clustering the data based on related themes, then discussed and assigned labels to each theme, and finally tagged the themes based on their relationship to social support (using the predefined codes listed above) \cite{ulrich_kj_2003}. Quantitative analyses of survey and app usage data were conducted using Qualtrics and Python, including descriptive analyses and paired sample t-tests (for assessing change between study touchpoints). The app usage data, survey data, and interview data were triangulated to add validity by assessing the app in multiple ways \cite{patton_qualitative_2002}.

\section{Findings}

\subsection{Participants}

Twenty-two eligible individuals completed Touchpoint 1 of the study, nineteen participants completed Touchpoints 1 and 2, and fifteen participants completed all three touchpoints. Participants who completed all three touchpoints ranged from age 17 to 24, with an average age of 21. Although seven participants under the age of 20 were consented at the start of the study (32\% of all consented participants), only four (27\% of all completing participants) completed all phases of the study. This indicates that age may have been a factor in whether participants completed the study. 47\% of participants were presently in foster care, and 53\% of participants were formerly in foster care. Participants who were formerly in care indicated that they had left foster care between the ages of 17 and 21, with an average age of 18. 43\% of participants had experience with group homes, 39\% had experience with foster homes, and 11\% had experience with kinship care. In terms of gender, 73\% identified as women and 27\% identified as men, with no participants identifying as non-binary or gender non-conforming. The participant pool was overrepresented in marginalized experiences across domains of sexuality and race, which may indicate that results speak to the perspectives of those who are most in need of support. 41\% of participants identified as LGBTQ+, compared to national estimates that 30\% of foster-involved youth identify as LGBTQ+ \cite{baams_lgbtq_2019,matarese_cuyahoga_2021,sandfort_experiences_2019}. In terms of race, 33\% identified as Black/African American and 27\% as Hispanic/Latinx, with only 33\% identifying as White/Caucasian (compared to 43\% of foster-involved youth identifying as White/Caucasian nationally) \cite{bureau_adoption_2023}. Participants who did not complete all three touchpoints were excluded from the subsequent analyses as they were not compensated for their contribution to the study. Of the participants who completed all three touchpoints, 5 (33\%) completed follow-up interviews as well.

\subsection{General trends}

Over the two-week intervention period, the fifteen participants contributed a total of 168 check-ins, 128 reactions, and 16 replies on the platform. On average, participants completed check-ins on 75.6\% of the days (adjusted for differing onboarding dates). On an average day, 52\% of participants (n = 7.8) checked the community feed page and 60\% (n = 9) checked the memories page (using view data from the last six days of the study). As described in the Methods section, compensation was based solely on completing surveys and daily check-ins, which suggests participants left reactions and replies and viewed their feed and memories out of an intrinsic interest in the app.

Overall, participants were generally satisfied with the experience of using the app. When asked how likely they would be to use the app in the future, 80\% (n = 12) of survey respondents said they would be Likely (47\%) or Extremely Likely (33\%) to do so. Additionally, 80\% (n = 12) of respondents said they would definitely recommend the app to current and former foster-involved youth. When asked what their overall star rating of the app would be (from 1 to 5), the average rating given was 4.4, with a minimum rating of 3.5 (given by 20\% of participants, n = 3) and a maximum rating of 5 (given by 47\% of participants, n = 7). When asked to rate how much each feature of the platform motivated them to keep using the app (1 = Not at all to 5 = A lot), participants indicated that they were especially motivated by reading other people’s check-ins (avg = 4.2, std dev = 1.3), selecting a feeling for their check-in post (avg = 4.1, std dev = 1.1), revisiting their past check-ins (avg = 4.0, std dev = 1.1), and writing highs and lows for their check-in posts (avg = 3.9, std dev = 1.6).

App usability received a score of 75.7 (std dev = 20.8) on our modified System Usability Scale (see Appendix for details). This put the app in roughly the 70 - 79th percentile range, as an average score is generally considered to be 68 \cite{noauthor_item_2018}. Looking at specific items from the scale, the majority of respondents thought the app was easy to use (86\%, n = 12), felt very confident using the app (86\%, n = 12), and felt that people would learn to use the app very quickly (87\%, n = 13). The main negative feedback was that 34\% of respondents (n = 5) found the app very cumbersome to use, although this was not corroborated in open-ended survey responses or follow-up interviews. 

\subsection{Emotional support}


The app appeared to provide participants with emotional support. On the post-intervention questionnaire, 67\% of participants (n = 10) said that the app helped them feel understood or less upset about something most days (47\%, n = 7) or every day (20\%, n = 3). These participants added that the app “helps [you] cope with everyday life” (P47, 20+ y/o, former foster) and “...helps with dealing with your feelings” (P38, 20+ y/o, former foster). While many participants mentioned feeling sad/alone (n = 5), anxious/scared (n = 4), or frustrated/angry (n = 4) in their check-ins, check-ins were most commonly tagged with the feeling “Calm” (22.6\%, n = 38), followed by “Relaxed” (13.1\%, n = 22), “Content” (12.5\%, n = 21), and “Excited” (11.3\%, n = 19). This may have been in part because using the app makes users feel better. 

Participants elaborated that reading other people’s check-ins and receiving replies or reactions on their check-ins helped them feel emotionally supported. When asked to rate whether features of the app helped them reflect on their feelings (1 = Not at all to 5 = A lot), participants generally felt that reading other people’s check-ins was helpful in the self-reflection process (avg = 3.8, std dev = 1.4). In follow-up interviews, participants spoke about how receiving supportive replies (n = 1), reactions (n = 2), or reading other people’s posts (n = 1) made them feel more understood or less upset. One such participant (P41, 20+ y/o, current foster) explained why reactions on their check-ins helped them feel better: “...if someone likes your check-in or something that you put up there, it's like, okay, they understand. They can relate.” 

Multiple participants tried to provide emotional support to other users, with 60\% (n = 9) reporting on the post-intervention questionnaire that they tried to help others feel understood or less upset on the app (13\% every day, 7\% most days, 13\% half the days, 27\% a few days). This seemed to take place through both reactions and replies. In an analysis of app check-ins and replies, three users responded to posts with empathetic responses, either relating to them or letting them know they care. In response to a check-in about an exciting job interview and a son that is very sick, 
one participant (P45, 20+ y/o, current foster) wrote: “Congrats on the job opportunity!! I hope your [son] feels all better very soon [green heart emoji].” In another instance, when a participant talked about being relaxed but bored at home, 
P50 (20+ y/o, former foster) related by saying “Home brings peace it’s just a little too much peace sometimes.” In follow-up interviews, two participants also talked about using reactions as a way to provide emotional support, with one participant (P37, 20+ y/o, former foster) saying that if they noticed another user was upset about something, they would react to the post to show that “I would probably feel that way too if I was going through that.”

Participants described how the app helped them reflect on and moderate their emotions. In interviews, two participants described the app's ability to promote self-reflection, with one (P51, 14-19 y/o, current foster) explaining that the app “gives you a chance to reflect on yourself more than you would if you didn’t have it.”
When asked to rate whether different features of the app helped them reflect on their feelings (1 = Not at all to 5 = A lot), participants indicated that writing check-ins (avg = 4.6, std dev = 0.8) and revisiting their past check-ins (avg = 4.7, std dev = 0.6) were helpful in the self-reflection process. Through the post-intervention questionnaire and follow-up interviews, seven participants elaborated on the ways that the check-in process helped them self-reflect and cope with emotions. 
For example, 
P44 (20+ y/o, former foster) said: “I wrote one day a grown-up woman was acting like a 2-year-old, but just didn't realize that I didn't make the situation any better by making fun of her… After I wrote it, when I actually looked at the post, I was like, damn, you could have done better too.” Participants' Self-Reflection Scale scores between app onboarding (Touchpoint 2, avg = 30.5\textpm4.8) and post-intervention questionnaire (Touchpoint 3, avg = 31.4\textpm7.5) trended upwards but did not change significantly.

Through follow-up interviews, three participants talked about wanting more ways to get emotional support on the app, sharing ideas like therapeutic games, in-app therapists that can provide ideas, or a feature where people can express their feelings through music. Four participants also specifically said they wanted the check-in form to include more feeling options or the ability to write in their own feelings. As P45 (20+ y/o, current foster) explained, “I think the feature for using a word to describe your feelings should be more inclusive because almost every time I was selecting one there wasn’t really something that fully described how I was feeling.”

\subsection{Social network support}

\begin{figure}[t]
    \centering
    \begin{minipage}[b]{0.44\textwidth}
        \includegraphics[width=\textwidth]{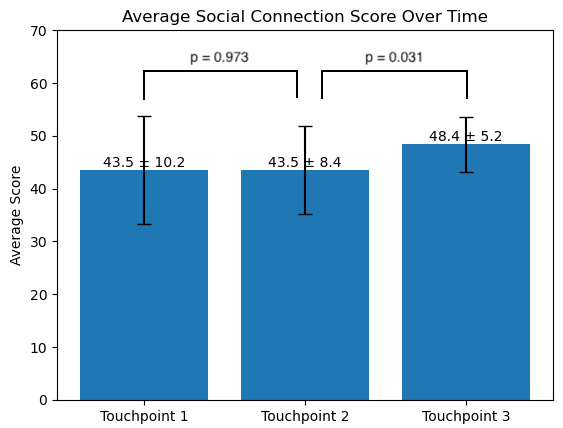}
        \caption{Average Social Connection Score over time}
        \Description{A histogram where each bar represents the average Social Connection Score at each study touchpoint.}
        \label{fig:socialconnection}
    \end{minipage}
    \hfill
    \begin{minipage}[b]{0.52\textwidth}
        \includegraphics[width=\textwidth]{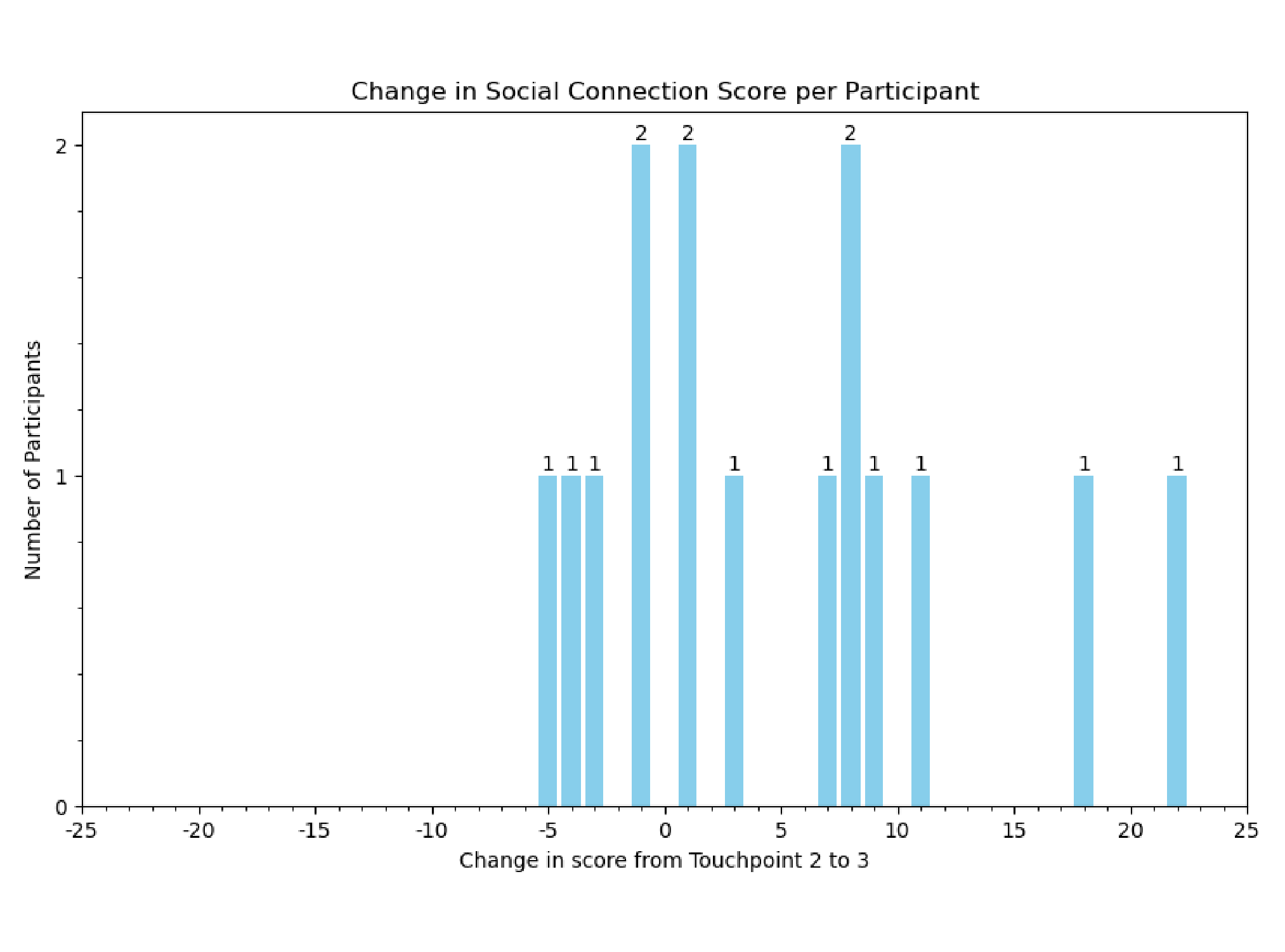}
        \vspace{-1cm}
        \caption{Change in Social Connection Score per participant}
        \Description{A histogram where each bar represents a participants' change in social connection score between Touchpoints 2 and 3.}
        \label{fig:socialconnectionperperson}
        \vspace{.5cm}
    \end{minipage}
\end{figure}

Across many data sources, the app appeared to provide participants with social network support. Participants showed a statistically significant increase (p = 0.031) in their Social Connection Scale score between app onboarding (Touchpoint 2) and the post-intervention questionnaire (Touchpoint 3) (Figure \ref{fig:socialconnection}), with 67\% (n = 10) experiencing a score increase (see Figure \ref{fig:socialconnectionperperson} for the distribution of score changes between Touchpoints 2 and 3). Additionally, on the post-questionnaire survey, 60\% (n = 9) of participants said that the app helped them feel less alone or more connected to others most days (40\%, n = 6) or every day (20\%, n = 3). For example, one of these participants elaborated, “You don't feel alone[, like you’re] the only one going through these things” (P44, 20+ y/o, former foster).  
When asked whether the app ever made them feel less alone or more connected to others during follow-up interviews, five participants responded affirmatively and pointed to reading others’ check-ins as the main reason why. One such participant (P38, 20+ y/o, former foster) explained that the app made them feel more connected because, “I read a lot of other people's posts… I'm reading somebody's stuff [who has] been through the same sort of [things].” 

Posts often shared common themes, which may be why participants felt a strong sense of social connectedness on the platform. From app data, we saw that all fifteen users wrote check-ins about school or work (working towards getting a job, trying to get to work/school on time, doing well in assignments, feeling tired because of work/school, and taking time off). Twelve users wrote about sleeping too little or too much (often to catch up on sleep), or generally being tired. Nine users shared highs and lows related to others in their community (children, friends, family, significant others). Nine users described experiences with food (either being hungry, full, or eating something they liked). Nine users talked about doing or needing to do daily life tasks like cleaning, dealing with taxes/bills, scheduling appointments, going shopping, and exercising. Eight users talked about experiencing physical discomfort (either being hurt or sick). Seven users wrote about being excited to go outside (to go shopping, walk in nature, or generally not be at home). Overall, these responses align with what current and former foster-involved individuals talk about on other social platforms (based on an analysis of a foster care-specific reddit channel), especially topics of education, physical health, employment, and family relationships \cite{fowler_fostering_2022}. 

Additionally, participants tried to help others feel less alone or more connected on the app, with 73\% (n = 11) of participants reporting on the post-intervention questionnaire that they tried to do this every day (7\%), most days (13\%), half the days (33\%), or a few days (20\%). From an analysis of app data, five participants used check-in comments as a way to connect with others over shared interests or challenges. In one instance of this, two participants (P38, 20+ y/o, former foster; P51, 14-19 y/o, current foster) connected over a television show they both liked. The participant who first mentioned the show in their check-in (P38) reflected in a follow-up interview that “talking about a show that we were watching… like the people in the show who we like” made them feel “understood.” 

Through the post-intervention questionnaire and follow-up interviews, six participants expressed interest in having more ways to interact with others on the app. Two of these participants (P40, 20+ y/o, current foster; P44, 20+ y/o, former foster) talked about wishing that they had received more reactions or replies. Both participants received some reactions but few or no replies. Two participants added that commenting and reacting alone did not feel like enough, with one (P45, 20+ y/o, current foster) saying, 
“It’s not very interactive, besides replying to others there not many other ways to connect." Three participants had suggestions of additions, with one stating, “Add video chat just in case we want to meet people” (P46, 20+ y/o, former foster), and the other two recommending incorporating things like games, app-generated daily/weekly questions, or more encouragement for users to ask each other questions. 

\subsection{Esteem support}

Participants appeared to acquire esteem support from the app, with 53\% (n = 8) of participants saying that the app helped them feel inspired or motivated to pursue their goals most days (20\%, n = 3) or every day (33\%, n = 5) on the post-intervention questionnaire. This may have been because the app provided users with a space to voluntarily share the successes and challenges they faced while working towards their goals via check-in posts. 

Indeed, many check-ins centered around the sharing of big and small challenges and accomplishments. One major theme was related to life goals like paying bills, cleaning, grocery shopping/cooking, and building or maintaining healthy relationships. Two participants talked about facing challenges in completing life goals, with one participant (P47, 20+ y/o, former foster) writing as their low: “I have a whole lot of organizing and cleaning to do. I need to stop procrastinating but can’t seem to find the courage.” Seven users shared moments when they were in the process of or had already accomplished life goals. 
For example, as their high one day, P3 (20+ y/o, former foster) said, “I got my food stamps card today so I did a big shopping trip. Also got my rugs and clothes nice and clean.” Another participant (P50, 20+ y/o, former foster) talked about working towards relationship-oriented goals, writing “I have released a toxic relationship” as their high, and “I have to learn how to live without him” as a low they were still working towards. 

Another theme was related to performance at school or work. Two participants shared lows related to not progressing in school or work tasks as much as they wanted. One participant (P38, 20+ y/o, former foster), for example, shared, “Didn’t complete any school work so I’m a little disappointed.” On the other hand, two users described accomplishments that they had at school. For example, as their high, one participant (P39, 14-19 y/o, current foster) wrote, “I got a A+ on my test and my cat came to me when I got home.” 
Interestingly, seven participants wrote specifically about working towards the goal of getting to school or work on time. Four participants centered check-ins on struggling to arrive on time or accomplish everything they wanted to do before they left the house, with lows like “I got to work very late” (P40, 20+ y/o, current foster) and “Didn’t have time to eat anything” (P41, 20+ y/o, current foster). Three users also expressed pride that they woke up and made it to school or work, with highs like, “I'm happy that I got to work on time” (P40, 20+ y/o, current foster).
Three participants talked about working towards getting a job (e.g., career courses, job interviews, job-related drug tests). For example, P50 (20+ y/o, former foster) talked about anticipating an upcoming job interview but dealing with weather-related obstacles, writing as their high, “I got a job interview but very nervous” and “it started to rain and couldn’t make it so I rescheduled” as their low, along with a feeling tag of “Disappointed.” 


Some participants appeared to actively try to provide esteem support to other users, with 53\% (n = 8) of participants reporting that they tried to help others feel inspired or motivated to pursue their goals on the app in the post-intervention questionnaire (13\% every day, 13\% most days, 7\% half the days, 20\% a few days). This occurred primarily through check-in messages and replies. In an analysis of app data, six participants wrote check-in messages aimed at encouraging others to continue working towards their goals. These included messages like: “Remember that even the smallest bit of progress is still progress” (P51, 14-19 y/o, current foster); “Everyone think Happy thoughts and always think good things like I can do this” (P39, 14-19 y/o, current foster). 
Additionally, two users wrote replies that tried to motivate others to overcome challenges they were facing. When P50 (20+ y/o, former foster) wrote about being nervous for an interview and being disappointed that they had to reschedule it because of bad weather, another participant (P44, 20+ y/o, former foster) provided esteem support by saying, “You got this girl. I am cheering you on and I know your going to get this job. Send you positive vibes.” In a follow-up interview, P44 shared that they had intentionally written an encouraging reply”: “I just told her to keep her head up because she was nervous. I said she was gonna get the job… I feel like it's always good to empower other people.” 

\subsection{Informational support}

Overall, the app did not seem to provide participants with substantial informational support. On the post-intervention questionnaire, only 27\% (n = 4) of participants said that the app helped them get useful advice or information most days (20\%, n = 3) or every day (7\%, n = 1). Additionally, when asked in follow-up interviews whether the app helped them get useful advice or information, five participants said that it did not. Four of these participants felt that the app was more focused on feelings and on self-reflection, with three explaining that it did not feel like there was an explicit place to ask a question. Three participants felt that they would be interested in asking questions if there was a clearer way to do so. One participant (P3, 20+ y/o, former foster) explained, “It was more feelings oriented… because there wasn't no section to just outright ask the question…” Indeed, an analysis of the app data showed that no participants added a question to their check-in (which they could have done within the message field). 

Interestingly, some participants still reported trying to provide informational support to others on the platform, although less frequently than other types of support. Amongst the 53\% (n = 8) of participants who reported trying to give useful advice or information on the app in the post-intervention questionnaire, none said they did this every day, 13\% (n = 2) said most days, 7\% (n = 1) said half the days, and 33\% (n = 5) said a few days. Looking at the app data, only two participants shared informational support in response to other users’ check-ins. For instance, when another participant talked about finding a mouse in their apartment, P51 (14-19 y/o, current foster) replied with a tip, “If u have peppermint oil or anything peppermint just spray (or put) around the apartment they hate it.” 

In follow-up interviews, five participants said that they never tried to give advice or share information on the app. Two said they did not feel like they had enough relevant information to be able to provide guidance to other users. Three participants explained that they did not feel like it was their place to provide guidance because other users were mostly sharing how they were feeling rather than asking for advice. Two participants added that they would have provided guidance if other users had asked specific questions. P37 (20+ y/o, former foster) summed this up, saying that they did not provide guidance because “I [didn’t] think [it was] appropriate for the setting depending on what people were talking about… like for me to just share a resource, it's probably something they already know about… [they] know how to how to get help, how to talk to somebody… if they had a question… then I'd be able to answer it.”

\subsection{Psychological safety}

The platform seemed to provide participants with some sense of psychological safety, or a freedom to express thoughts and feelings without consequences \cite{rogers_toward_1954}. Eight participants highlighted anonymity as a core reason why they felt comfortable writing in the app. Four participants felt that anonymity made it easier for them to interact with others on the platform, with one participant (P38, 20+ y/o, former foster) saying, “I love how we got to interact with people without judging a book by its cover.” Four participants also talked about how anonymity made it easier for them to express themselves in their check-ins, with one (P41, 20+ y/o, current foster) explaining, “What I really like [is that] it's anonymous, only because I feel like it was easier to do your highs and your lows… [it’s] kind of private, but I'm still expressing how I feel.” On the other hand, two participants expressed interest in being able to privately communicate with users they relate to, because it would make them feel more comfortable sharing personal information or make them feel less worried about being judged by others. 

As evidence of their comfort on the platform, participants seemed to share intimate challenges or feelings they were struggling with in their check-in posts. In their lows, five participants wrote about feeling sad or alone, with posts like, “Som[times] I feel single[d] out by the people I'm consistently around” (P47, 20+ y/o, former foster) and “I feel Sad because my aunt died” (P43, 20+ y/o, current foster). 
Four participants shared that they were feeling anxious or scared, for example, writing, 
“I’m scared to trust again” (P50, 20+ y/o, former foster). Four participants also described moments when they were feeling frustrated or angry, like one (P50, 20+ y/o, former foster) who wrote, “Child custody is a lot” and “still mad I’m going through this with my first kid.” Four participants also wrote check-ins about personal struggles they faced with family or in romantic relationships, with one participant (P42, 20+ y/o, current foster) sharing lows like “had to talk to my mum” and “almost [losing] a friend.” 

Additionally, an analysis of app data suggested that no check-ins or replies violated the rules set up to protect users on the platform. Only 4.6\% of check-ins (n = 7) and zero replies were auto-flagged by the content moderation system as containing potentially inappropriate content, with zero found to actually be in violation of community guidelines upon moderator review. Content was auto-flagged for use of flagged keywords (including "drug test" and "binging" in appropriate contexts, and as a misspelling, "breasts"), and use of numbers or strings with periods in the middle (categorized as potentially sharing identifiable information). 
Furthermore, no check-ins or replies were flagged by users as inappropriate and
in a retroactive analysis of all the data, the research team determined that none violated the community guidelines. 

\section{Discussion}

In this study, we designed and evaluated Sphere, the first digital platform designed and shown to provide emotional, social network, and esteem support to foster-involved youth. This system fills important gaps in care for foster-involved youth \cite{okpych_memo_2018,herman_trauma_2015,kawachi_social_2001,cohen_stress_1985} and helps provide social support to heal from past traumas and develop resilience to future stressors \cite{herman_trauma_2015,kawachi_social_2001,cohen_stress_1985}.

Knowing of no prior studies which investigated the design affordances that facilitate online social support for foster-involved youth, we have drawn from work in other contexts for Sphere's design. In particular, we have integrated the following design principles to facilitate social, especially emotional, social network and esteem, support: 1) creating online spaces that connect individuals with common identities and experiences \cite{prescott_online_2020, naslund_future_2016,zhang_separate_2022,andalibi_considerations_2021,saksono_evaluating_2023,fang_matching_2022}; 2) supporting users in sharing about personal experiences \cite{progga_just_2023, choudhury_mental_2014,huh-yoo_help_2023}, aided by anonymity \cite{andalibi_social_2018,choudhury_mental_2014}; 3) focusing on preventing harmful actions, rather than restricting access to online platforms or just reactively supporting victims \cite{badillo-urquiola_risk_2019, badillo-urquiola_towards_2024,agha_strike_2023}. This, combined with our iterative co-design process with youth who have lived-experience, results in a first-of-its-kind digital platform specifically designed to provide social support to foster-involved youth, which 1) is only open to individuals with foster care experience (rather than being public); 2) scaffolds self-disclosure by centering it around balanced, self-reflective check-ins (rather than free-form posting); and 3) proactively promotes safety through anonymity and immediate removal of potentially triggering or harmful content (rather than only removing posts once they've been shown to violate platform rules). Analyzing how foster-involved youth interact with this new platform, we extend prior research by comparing the results with findings from other online platform contexts and highlighting the design decisions that appear to facilitate online social support for foster-involved youth. 

\subsection{Uniquely Fostering Emotional, Social Network and Esteem Support}

Participants primarily reported gaining emotional, social network and esteem support through the platform, in contrast to studies in other social platform contexts that highlight informational support as equally (if not more) prevalent than other types of social support \cite{huh-yoo_help_2023, grimes_eatwell_2008, zhang_separate_2022}. This may have been because Sphere encouraged youth to share personal stories about a variety of topics rather than asking for verified and focused information \cite{nakikj_park_2017}. Given that some participants still tried to give useful advice or information on the app or were open to doing so if there had been specific questions they could answer, we believe that peer-to-peer platforms could successfully share informational support for foster-involved youth if specifically designed for this purpose.

Interestingly, while prior research generally focuses on the supportive value of interacting with similar others online \cite{prescott_online_2020, naslund_future_2016,zhang_separate_2022,andalibi_considerations_2021,saksono_evaluating_2023,fang_matching_2022} and sharing personal experiences as a way to invite social support or support another user directly \cite{progga_just_2023, choudhury_mental_2014,huh-yoo_help_2023}, our study found that simply reading about others' similar experiences was an important way that youth gained emotional and social network support. While it is possible that youth would have emphasized the value of direct interaction more if there had been more replies exchanged on the platform, the fact that the vast majority of participants (80\%, n = 12) reported wanting to use the app outside of the study hints that the benefits of reading others' were substantial on their own. In contrast to prior works' concern that reading others' experiences could trigger unwanted negative feelings for users \cite{gold_online_2022, barta_similar_2023,randazzo_if_2023}, no participants in our study reported experiencing negative emotions from seeing others’ posts. This may have been because the check-in requirement to share both a "High" and a "Low" prevented overly negative or positive content (which could bring up hopelessness, inferiority, or unwanted associations), because the community norms did not allow discussion of some potentially triggering topics, or because youths' shared identities and experiences promoted socially supportive comparisons \cite{barta_similar_2023}. 

The reflective check-in template also appeared to support youth in reflecting on and moderating their emotions. Youth suggested that this happened both in the moment they wrote their post (allowing them to reflect on how they had been feeling that day) and in looking back at past posts (providing opportunities to reflect on their emotion regulation skills). This in-the-moment and retroactive reflection has been shown to be an added "self-support" benefit of self-disclosure in other online platform contexts as well \cite{luo_self-disclosure_2020, karizat_i_2023}. However, unlike the findings of previous research on social media users who post weight-related content \cite{karizat_i_2023}, no youth in our study described having unwanted negative thoughts or feelings based on their self-disclosures on the platform. This may have been due to the requirement to share both a positive and a negative aspect of one's day, which could have encouraged youth to remain balanced in their reflections, neither focusing too much on current challenges nor on positive things they no longer have.

Replicating the findings of work focused on public reddit communities \cite{andalibi_social_2018,choudhury_mental_2014}, anonymity seemed to help participants feel safe sharing their feelings and experiences on the platform, primarily through their check-in posts. However, despite feeling comfortable self-disclosing in check-in posts and finding resonance in other youths' posts, participants appeared less comfortable writing replies. Some participants expressed hesitation leaving public supportive replies because they were worried about how it would be received by the original poster and others in the community, a concern echoed in other sensitive online community contexts (for example, in discussions of traumatic experiences and life-threatening health conditions) \cite{randazzo_if_2023, smith_thoughts_2023}.

While studies highlight that there are many risks to foster youth engaging in online platforms \cite{gustavsson_positive_2015, badillo-urquiola_risk_2019}, no participants in this study reported encountering content that made them feel unsafe or uncomfortable. Additionally, a retroactive analysis of all app data did not find any instances of behavior that violated any community guidelines, which were defined to prevent potentially triggering or harmful content. This may have been because of a variety of factors, including the definitions and clear presentation of the community guidelines, the platform's restriction to individuals with verified lived experience with the foster care system, or the study’s small sample size. Larger studies should be conducted to determine whether these risks manifest when similar social support platforms are launched at scale with current and former foster-involved youth, and which safety mechanisms best protect users.


Based on the evaluation results, we present recommendations, below, for designing internet-connected platforms that engage foster-involved youth in socially supportive ways.

\subsection{Recommendation \#1: Create systems of self-reflection and subsequent community sharing}

Digital platforms can give foster-involved youth a place to reflect on their emotions and externalize their thoughts without the requirement of sharing with others. By starting with a self-reflection activity and then providing the option to share publicly, youth may find it easier to disclose personal experiences on the platform. In our study, 85\% of posts (n = 143) were made public even though this step was not necessary for saving posts to the private “Memories” archive or for study compensation. This self-disclosure can help youth receive social support from others, particularly if negative self-disclosure is solicited (as in Sphere, where a high and a low were required) \cite{huh-yoo_help_2023,luo_self-disclosure_2020,yang_channel_2019}. 

Additionally, youth can benefit from reading about the experiences of other foster-involved youth. In this study, youth highlighted that reading about what others were going through (especially seeing similarities in challenges and goals) made them feel less alone and more connected (social network support). By having youth share personal experiences first (unlike a traditional peer support platform that focuses on having youth ask questions for others to respond to), this design may be especially helpful for youth who are hesitant to reach out to others online.

\subsection{Recommendation \#2: Help youth send supportive messages to one another}

We recommend making it \textit{easy} for foster-involved youth to respond supportively to the personal experiences disclosed by their peers. In our project, youth talked about how receiving replies or reactions on their check-in posts helped them feel more understood and less upset (emotional support). However, we saw few youth use the reply feature of our app, perhaps because they did not know how to adequately help \cite{huh-yoo_help_2023} or were worried that others would judge or harass them based on what they shared in their reply \cite{randazzo_if_2023, smith_thoughts_2023}. We believe that designs should do more to help youth feel comfortable and equipped to reply to others. Allowing anonymous replies and providing templated responses or expressive stickers, for example, may reduce fear of judgment and help youth who have difficulty with writing to succeed in expressing support \cite{jin_understanding_2023, zhang_separate_2022, smith_thoughts_2023}. 

Interestingly, the findings of this study suggest that youth may have an easier time sharing motivational messages (esteem support) with the community as a whole. This may be because they have a desire to support others but do not always feel equipped to support individuals with specific situations. Providing more scaffolding to help youth craft supportive messages (such as the templated response concept presented above) or matching youth to others whose situations they can relate to may help make it easier for youth to provide more tailored support to their peers. 

To account for the possibility that young people are hesitant to engage with one another directly even with additional scaffolding, we suggest that designs be centered around constructive self-reflection and personal storytelling, as this appears to come naturally to individuals with lived experience \cite{fowler_policy_2023} and has multiple benefits that are not tied to user interaction: self-support for the poster through reflecting on past challenges they've overcome \cite{karizat_i_2023} and social support for other users through learning about their relatable experiences \cite{grimes_eatwell_2008}.

\subsection{Recommendation \#3: Design for informational support separately from emotional support}

We suggest that designers consider informational support separately from emotional, esteem, and social network support when designing digital tools for foster-involved youth. In our research, youth expressed a desire for clearer ways to find and share resources on the app (informational support); however, even when the research team seeded two check-in posts that requested informational support using the “message” feature, no users requested informational support themselves. This may have been due to the features we implemented to encourage emotional, esteem, and social network support, including the focus on self-reflection and experience sharing (rather than a focus on asking questions), anonymity, and not dividing the community by region. Anonymity, for example, helped create a psychologically safe environment in which youth felt comfortable sharing personal information (especially disclosing challenges or negative emotions) and were less worried about being judged by others. This is supported by existing literature, which describes how anonymity makes it more comfortable for users to share personal stories in support-seeking posts and support-providing comments \cite{choudhury_mental_2014,andalibi_social_2018}. The fact that our emotion-focused platform was unable to provide adequate informational support is line with research in other contexts that suggests that creating spaces for informational support and emotional support have different associated design decisions \cite{nakikj_park_2017}. Furthermore, given that mental health social platform comments containing informational support tend to be more negatively received by readers, we suggest that designers interested in fostering informational support for foster-involved youth think carefully about how to support young people in sharing information that their peers find useful and respond positively to \cite{kim_supporters_2023}. 
		
\section{Limitations and Future Work}

This work has several limitations. The study was conducted with a small group of current and former foster-involved youth in Massachusetts and California, 
and had a skewed gender breakdown, with 73\% identifying as women and 23\% as men, compared to the national breakdown of foster-involved youth, which is roughly 51\% male and 49\% female \cite{bureau_adoption_2023}. Thus, while the findings shed light on key perspectives and behavior of some foster-involved youth, they may not generalize to a broader population. We advocate that future studies investigate how the findings and design recommendations presented in this study translate to larger, more diverse populations. Additionally, while the sample originally included seven younger youth (between ages 17 and 19), only four of these completed the study and only one took part in a follow-up interview. Further research should try to include more participants between the ages of 16 and 19 and identify ways to keep this population engaged in the research process. Because the current study limited participants’ use of the app to two weeks, a longitudinal study is also needed to examine how use is impacted over a greater length of time. Additionally, because we incentivized participants to check in daily on the app, it is unclear whether the app would retain users in the wild. Future research should investigate user retention and engagement on the platform by decoupling compensation from participants’ app use. While we required users to share both a high and a low to encourage socially supportive interactions on the app, this could have prompted some users to recall negative experiences they would otherwise not have focused on. Future work should study whether this format has the potential to worsen mood for any users, and identify mitigating design features. Lastly, to keep the platform safe and comfortable, we prohibited discussions of suicide, self-harm, abuse and disordered eating, directing participants to use emergency support services if they needed support related to these topics. Although no participants attempted to discuss this content on the platform, this restriction may have kept participants from getting support with important and urgent challenges. Future interventions should explore content moderation strategies that allow for these topics to be discussed in ways that ensure participants get the support they need and are not exposed to triggering content.

\section{Conclusion}

This work explored if and how internet-connected technology can be used to provide social support to foster-involved youth. We involved lived-experience experts in every stage of development as we designed, built, and evaluated the first digital tool that aims specifically to increase social support for foster-involved youth. The novel tool we co-created, Sphere, provides an interface that allows youth to complete reflective check-ins and react and respond to other users’ check-ins in supportive ways. Through a study evaluating the usage and effects of this platform, we gathered qualitative and quantitative evidence to suggest that this type of system is able to provide youth with emotional, esteem, and social network support. Based on participants’ experiences with the app, we posit that digital platforms can increase emotional, esteem, and social network support for foster-involved youth by creating systems for them to privately reflect on their experiences and anonymously share their reflections and supportive messages with other youth who have had similar experiences. 

\begin{acks}
We thank our reviewers for their detailed and constructive feedback on this paper. We also express our gratitude to our participants and community organization partners, whose generous and thoughtful involvement was vital for every stage of this project. Lastly, thank you to the Shah Family Foundation and our other donors, who have made this project possible. 
\end{acks}

\bibliographystyle{ACM-Reference-Format}
\bibliography{sample-base}

\end{document}